\documentclass[tightenlines,aps,floatfix,preprint,
nofootinbib]{revtex4}
\usepackage{float,axodraw,psfig}

\newcommand{\ba}{\begin{eqnarray}}
\newcommand{\ea}{\end{eqnarray}}

\newcommand{\ep}{\epsilon}

\newcommand{\be}{\begin{equation}}
\newcommand{\ee}{\end{equation}}

\begin{document}

\title{
\hfill{\normalsize\vbox{%
\hbox{\rm UH-511-1069-05}
}}
\vspace*{0.5cm}

The electron energy spectrum in muon decay through ${\cal O}(\alpha^2)$
}

\author{
Charalampos Anastasiou\thanks{e-mail:babis@phys.ethz.ch}} 
\affiliation{
          Institute for Theoretical Physics,\\ 
          ETH, 8093 Zurich, Switzerland}
\author{Kirill Melnikov
        \thanks{e-mail: kirill@phys.hawaii.edu}}
\affiliation{Department of Physics and Astronomy,
          University of Hawaii,\\ 2505 Correa Rd., Honolulu, Hawaii 96822}  
\author{Frank Petriello\thanks{frankp@pha.jhu.edu}}
\affiliation{
Department of Physics, Johns Hopkins University, \\
3400 North Charles St., Baltimore, MD 21218
}

\bigskip

\begin{abstract}

We compute the complete ${\cal O}(\alpha^2)$ QED corrections  
to the electron energy spectrum in unpolarized muon decay, including the full 
dependence on the electron mass. Our calculation reduces the theoretical uncertainty on the electron energy 
spectrum well below $10^{-4}$, the precision anticipated 
by the TWIST experiment at TRIUMF, which is currently performing this measurement. For  this calculation, we extend techniques we 
have recently developed for performing 
next-to-next-to-leading order computations to handle 
the decay spectra of massive particles.  Such an extension 
enables further applications to 
precision predictions for $b$, $t$, and Higgs differential decay rates.  

\end{abstract}

\maketitle

\section{Introduction}

The decay of a muon into an electron and a pair of neutrinos, 
$\mu \rightarrow e \nu_{\mu} \bar{\nu}_e$, 
occupies an important role in particle physics.  
The measurement of the muon lifetime \cite{Bardin:1984ie} leads to  
the most accurate  determination of the Fermi coupling constant, $G_F$.  
The muon  anomalous magnetic moment 
is one of the most precisely measured quantities in nature
\cite{Bennett:2002jb,Brown:2001mg}, 
and provides important constraints 
on physics beyond the Standard Model (SM) \cite{czar}.  Searches for lepton flavor-violating decays of the muon, such 
as $\mu \rightarrow e \gamma$ and $\mu \rightarrow eee$, constrain the flavor sector of many SM extensions~\cite{kuno}.

The calculations of radiative corrections to muon decay have a long and 
storied history \cite{kinoshita}.  The one-loop 
QED corrections were first performed within the Fermi theory of weak interactions in the 1950s~\cite{firstcalcs}.  
The cancellation of mass singular terms such as ${\rm ln}(m_{\mu}/m_{e})$ in the total rate, but not in 
distributions such as the electron energy spectrum, led to the development of 
the Kinoshita-Lee-Nauenberg theorem, which 
explains how to construct ``infrared-safe'' observables in quantum field theory where such effects 
cancel~\cite{KLN}.  The calculation of the full one-loop corrections in the ${\rm SU(2)} \times {\rm U(1)}$ theory of the 
electroweak interactions was one of the first such computations performed~\cite{EW1loop}.  The full two-loop 
corrections to the muon lifetime in the Fermi model, needed for a precision determination of $G_F$, were 
completed several years ago~\cite{QED2loop}; recently, the two-loop results 
in the full electroweak theory were obtained \cite{twoloopew}.

Muon decay continues to be of interest in particle physics.  The TWIST experiment at TRIUMF~\cite{TWIST} 
measures the electron energy and angular distributions in polarized muon decay; 
the first results were recently reported in \cite{twistik}.
It is anticipated that TWIST will eventually measure 
the Michel parameters~\cite{Michel}, which describe muon decay for the most 
general form of the four-fermion interaction, to a precision of $\approx 10^{-4}$.  This significantly increases 
the sensitivity of muon decay to deviations arising from new physics.  For example, the lower bound on the 
mass of the $W_R$ in the manifest left-right symmetric model is improved from $M_{W_R} > 400$ GeV to 
$M_{W_R} > 800$ GeV, competitive with limits coming from the Tevatron, while the bounds on the left-right 
mixing parameter $\zeta$ are improved by nearly an order of magnitude~\cite{kuno}.  
Such precision requires a careful consideration of the higher order corrections.  As noted above, the 
radiative corrections to quantities such as the electron energy distribution contain large logarithms of 
the form ${\rm ln}(m_{\mu}/m_{e})$, which enhances their effect. The presence of mass singularities makes it 
impossible to compute the radiative corrections to the electron energy spectrum by neglecting the mass of the electron from 
the very beginning, the approximation that has been used successfully in the 
calculation of QED corrections to the muon lifetime \cite{QED2loop}. This 
feature makes the calculation of the ${\cal O}(\alpha^2)$ 
corrections to the spectrum a challenging problem that has defied solution for many years.

It was realized recently that the logarithmically enhanced parts of the second order QED corrections 
can be computed using the factorization of mass singularities traditionally 
discussed in the context of QCD. In this way, the two-loop corrections with a double 
logarithmic enhancement, ${\cal O}(\alpha^2 {\rm ln}^2(m_{\mu}/m_{e}))$, were calculated in~\cite{Arbuzov:2002pp}, 
and the singly-enhanced ${\cal O}(\alpha^2 {\rm ln}(m_{\mu}/m_{e}))$ terms 
were computed in~\cite{Arbuzov:2002cn}.  
At the midpoint of the electron energy spectrum, the sizes of these two terms are respectively 
$-7 \times 10^{-4}$ and $3 \times 10^{-4}$.  There are two interesting features of these results.  The first is 
that the corrections are larger than the anticipated experimental precision, $10^{-4}$.  The second is that 
the single-logarithmic terms are not a full factor of ${\rm ln}(m_{\mu}/m_{e}) \approx 5$ times smaller 
than the double-logarithmic terms, indicating that the naive power-counting based on the size of this 
logarithm might not hold.  Both of these facts render a full calculation of the ${\cal O}(\alpha^2)$ 
corrections desirable.

In this paper we compute
the ${\cal O}(\alpha^2)$ QED corrections to the electron energy 
spectrum in muon decay.  The full dependence on the electron mass is 
retained.  We use a method of performing next-to-next-to-leading-order 
(NNLO) calculations developed by us in a recent 
series of papers~\cite{secdecomp}.  Our technique features an 
automated extraction and numerical cancellation 
of divergences which appear as poles 
in the dimensional regularization parameter $\ep = (4-d)/2$.  In muon 
decay, ultraviolet divergences and 
divergences arising from soft photon emissions 
appear as $1/ \ep$ poles, while emission of photons along the 
electron direction is regulated by the finite mass 
of the electron and leads 
to logarithms of the ratio of the muon to electron mass, 
${\rm ln}(m_{\mu}/m_{e})$. From the technical point of view, the fact 
that the electron mass plays the role of the collinear regulator leads 
to some differences as compared to calculations with only massless 
particles.  Having masses as regulators reduces the complexity of the 
analytic structures which must be integrated over multi-particle phase-spaces 
or over virtual loop momenta.  However, issues of numerical stability appear since multi-dimensional
integrals are regulated by $(m_e/m_\mu)^2 \sim 10^{-5}$. We find that 
the presence of a mass regulator significantly simplifies the treatment 
of real emission processes of the type $\mu \to \nu_\mu \bar \nu_e e + \gamma\gamma$;
however, the computation of virtual corrections 
becomes more complex, compared to a purely massless case. 
We describe these and other technical 
issues in detail in the main body of the paper.

Many other physics 
applications require computations of higher 
order corrections to the decay spectra of massive particles.
 For example, the structure of the 
${\cal O}(\alpha^2)$ corrections 
to muon decay is identical to the ${\cal O}(\alpha_{s}^2)$ QCD corrections to 
semi-leptonic $b \rightarrow u$ 
and $b \rightarrow c$ transitions, which are used to extract 
the CKM matrix elements $|V_{ub}|$ and $|V_{cb}|$, 
the $b$-quark mass and other important parameters of Heavy Quark Effective Theory \cite{bphys}.
The calculation of QED radiative corrections to the electron energy spectrum 
discussed in this paper  can be easily 
extended to obtain differential results for semi-leptonic $b$-decays at NNLO.  In fact, some of the 
technical issues become simpler for $b$-decays, particularly $b \rightarrow c$.  In this case, collinear 
singularities are regulated by the factor 
$(m_c /m_b)^2 \approx 4 \times 10^{-2}$, rather 
than $(m_e /m_{\mu})^2 \approx  10^{-5}$, 
leading to more stable numerics.  Precise predictions for heavy particle decay spectra will also 
be important for measurements at both the LHC and a future linear collider.  Both experiments will 
search for anomalous top quark couplings through final-state distributions in its decay $t \rightarrow bW$, 
and will determine the CP properties of any scalar boson discovered through angular properties of 
such decays as $\phi \rightarrow ZZ,WW,f\bar{f}$~\cite{tophiggs}.  The techniques required to analyze higher-order 
corrections to these decay modes are very similar to those presented here.
 
This paper is organized as follows. In the next Section we introduce our 
notation and discuss general aspects of the computation of the electron 
energy spectrum.
In Section~\ref{sect.nnlo} we describe our computation of the NNLO QED corrections. 
In Section~\ref{sect.res} we discuss our results. Finally, we present our conclusions.

\section{Notation and setup}

\begin{figure}[htp]
\begin{center}
\begin{picture}(300,50)(0,0)
\SetColor{Blue}
\SetWidth{1.0}
\Line(0,35)(35,0)
\Line(35,0)(70,35)
\SetColor{Red}
\Vertex(35,0){3}
\SetColor{Black}
\put(0,40){$\mu$}
\put(70,40){$e$}

\SetColor{Blue}
\Line(110,35)(145,0)
\Line(145,0)(180,35)
\SetColor{Red}
\Vertex(145,0){3}
\SetColor{Black}
\put(110,40){$\mu$}
\put(180,40){$e$}
\Photon(120,25)(170,25){2}{4}

\SetColor{Blue}
\Line(220,35)(255,0)
\Line(255,0)(290,35)
\SetColor{Red}
\Vertex(255,0){3}
\SetColor{Black}
\put(220,40){$\mu$}
\put(290,40){$e$}
\Photon(230,25)(255,35){2}{4}

\end{picture}
\end{center}
\caption{\label{effdiags} A sample of LO and NLO diagrams which appear for the effective $\mu-e$ current after the neutrinos are 
integrated out.  The factor to be associated with the effective $\mu-e$ vertex after squaring the matrix elements is 
given in Eq.~\ref{neutproj}.}
\end{figure}
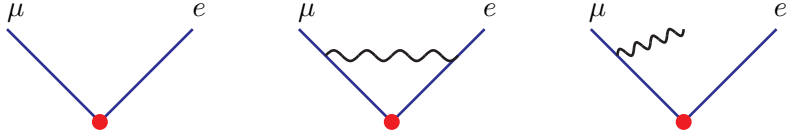

We discuss in this Section some basic notation needed to describe muon decay.  We begin with the Lagrangian 
\begin{equation}
{\cal L} = {\cal L}_{QED} + {\cal L}_{F}.
\label{lagtot}
\end{equation}
${\cal L}_{QED}$ contains the kinetic terms for the fermions and photons, along with the 
QED interactions,
\begin{equation}
{\cal L}_{QED} = -\frac{1}{4}F_{\mu\nu}F^{\mu\nu} + \sum_{f} \bar{\psi}_f \left[ i\not\! D -m_f \right] \psi_f,
\label{lagqed}
\end{equation}
while ${\cal L}_{F}$ contains the Fermi interaction,
\begin{equation}
{\cal L}_{F} = -2\sqrt{2}G_F \left[\bar{\psi}_{\nu_{\mu}}\gamma^{\rho} P_L \psi_{\mu} \right]
              \left[ \bar \psi_{e} \gamma_{\rho} P_L \psi_{\nu_{e}} \right].
\label{lagfermi}
\end{equation}
Here, $P_L = (1-\gamma_5)/2$ is the usual 
left-handed projection operator.  The Fermi Lagrangian can be 
Fierz rearranged into 
the following form:
\begin{equation}
{\cal L}_{F} \rightarrow -2\sqrt{2}G_F \left[\bar{\psi}_{e}\gamma^{\rho} P_L \psi_{\mu} \right]
              \left[\bar{\psi}_{\nu_{\mu}}\gamma^{\rho} P_L \psi_{\nu_e} \right].
\label{lagfierz}
\end{equation}
The QED corrections to this Lagrangian are finite to all orders in $\alpha$~\cite{fermiqed} after the 
fermion mass renormalization is included.  Since the 
QED corrections do not affect the neutrino part of this Lagrangian, and experiments do not 
probe properties of the neutrinos, they can be integrated out to produce an effective 
$\mu-e$ current.  We demonstrate this here.  We first note that since QED interactions only affect 
the leftmost fermion bilinear of Eq.~\ref{lagfierz}, we can write the squared matrix element for the 
process $\mu \rightarrow e \nu_e \nu_{\mu}+X$ as 
\begin{equation}
|{\cal M}|^2 = |{\cal M}^{\rho\sigma}_{\mu \rightarrow e+X}|^2 \times {\rm Tr}\left[\not\!p_{\nu_e}\gamma_{\rho}
  \not\!p_{\nu_{\mu}}\gamma_{\sigma}P_L\right].
\label{eq.55}
\end{equation}
${\cal M}^{\rho\sigma}_{\mu \rightarrow e+X}$ denotes the matrix element formed from the leftmost bilinear of 
Eq.~\ref{lagfierz} together with any QED corrections.  
Eq.(\ref{eq.55}) must be integrated over the appropriate phase-space 
to obtain the electron energy spectrum:
\begin{eqnarray}
\frac{d\Gamma}{dx}&=& \int [d \Pi_{\mu \rightarrow e\nu\nu+X}] |{\cal M}|^2 \nonumber \\ 
   &=& \int dp_{\rm nt}^2 
  \int [d \Pi_{\mu \rightarrow ep_{\rm nt}+X}]  
|{\cal M}^{\rho\sigma}_{\mu \rightarrow e+X}|^2 
  \times \int [d \Pi_{p_{\rm nt} \rightarrow \nu_{e}\nu_{\mu}}] {\rm Tr}\left[\not\!p_{\nu_e}\gamma_{\rho}
  \not\!p_{\nu_{\mu}}\gamma_{\sigma}P_L\right],
\end{eqnarray}
where $x = 2E/m_\mu$ and $E$ is the electron energy.
In the second line, we have partitioned the phase-space so that the muon first decays into an electron, 
additional radiation denoted by $X$, and 
a massive state with momentum
\begin{equation}
p_{\rm nt} = p_{\mu}-p_{e} -p_{X}.
\label{neutmom}
\end{equation}    
This massive state then decays into the muon and electron neutrinos.  The neutrino portion of the phase-space can 
be integrated over to obtain
\begin{equation}
\int [d \Pi_{p_{\rm nt} \rightarrow \nu_{e}\nu_{\mu}}] {\rm Tr}\left[\not\!p_{\nu_e}\gamma_{\rho}
  \not\!p_{\nu_{\mu}}\gamma_{\sigma}P_L\right] = T_{\rho\sigma}^{\rm nt},
\end{equation}
with
\begin{equation}
T_{\rho\sigma}^{\rm nt} = -\frac{\pi p_{\rm nt}^2}{3(2\pi)^{d-1}} 
\left[ g_{\rho\sigma} -\frac{ p_{\rm nt, \rho}
           p_{\rm nt, \sigma}}{p_{\rm nt}^2}
 \right],
\label{neutproj}
\end{equation}
giving the expression for the decay spectrum
\begin{equation}
\frac{d\Gamma}{dx} =  \int dp_{\rm nt}^2 \int [d \Pi_{\mu \rightarrow ep_{\rm nt}+X}]  
 |{\cal M}^{\rho\sigma}_{\mu \rightarrow e+X}|^2 T_{\rho\sigma}^{\rm nt}.
\label{eq.not.1}
\end{equation}
Since, up to an overall numerical factor, $T_{\rho\sigma}^{\rm nt}$ coincides with the 
polarization density matrix of a vector boson with mass $p_{\rm nt}^2$, 
we can interpret Eq.~\ref{eq.not.1} as the emission of such a boson in the 
$\mu-e$ transition.
A sample of the diagrams that appear in this description are shown in Fig.~\ref{effdiags}. 

Since we are regulating divergences in dimensional regularization, we briefly discuss our treatment of the 
$\gamma_5$ which appears in $P_L$ in Eq.~\ref{lagfierz}.  Our conclusion is that a naive anti-commuting 
$\gamma_5$ can be used, and fermion traces containing an odd number of $\gamma_5$ matrices do not contribute.  
This follows from the observation that the tensor $T^{\rho\sigma}_{\rm nt}$ is symmetric under 
$\rho \leftrightarrow \sigma$; a contribution containing an odd number of $\gamma_5$ matrices will produce 
a form factor containing the completely anti-symmetric Levi-Cevita tensor and will vanish when 
contracted with $T^{\rho\sigma}_{\rm nt}$.

We now discuss the form in which we will present our results.  
We parameterize the electron energy spectrum  with the 
variable $x=2 E/m_{\mu}$ which lies in the range 
\begin{equation}
\frac{2 m_e}{m_{\mu}} \leq x \leq 1+\frac{m_{e}^{2}}{m_{\mu}^2}.
\label{xrange}
\end{equation}
We write the differential decay rate as 
\begin{equation}
\frac{d \Gamma}{dx} = \frac{G_{F}^{2}m_{\mu}^{5}}{192 \pi^3} \sum_{n=0} \left(\frac{\alpha}{\pi}\right)^{n} f^{(n)}(x).
\label{decaydef}
\end{equation}
The LO and NLO results were computed in \cite{firstcalcs}; they can be obtained in a 
convenient form in~\cite{QED2loop,arbuzov1l}.  
The logarithmically enhanced contributions to $f^{(2)}(x)$ 
can be obtained from~\cite{Arbuzov:2002pp,Arbuzov:2002cn}.  The calculation of $f^{(2)}(x)$ beyond 
the logarithmic approximation, and including the full dependence on the electron mass, is the main 
subject of this paper.

\section{NNLO corrections}
\label{sect.nnlo}

In this Section we discuss our computation of the NNLO corrections to 
the electron energy spectrum. We give a brief overview 
of the technical aspects of the calculation and then describe in detail 
the computation of double real emission, one-loop virtual 
corrections to a single photon emission, and two-loop virtual corrections.

\subsection{Overview of NNLO corrections}

We first present a brief overview of the various components of the NNLO corrections.  The differential decay rate 
contains a sum over several distinct components,
\begin{equation}
\frac{d\Gamma}{dx} = \sum_{Y} \frac{d\Gamma_{Y}}{dx},
\end{equation}
where each $d\Gamma_{Y}/dx$ is separately divergent and must be combined with the other components to 
produce a finite result.  Our method of calculation follows 
the technique outlined in~\cite{secdecomp}.  We regulate both infrared and ultraviolet divergences in dimensional regularization, setting the 
space-time dimension $d=4-2\ep$, and produce an expansion 
\begin{equation}
  \frac{d\Gamma_{Y}}{dx} = \sum_{i=2}^{0} \frac{A^{Y}_{i}(p_{\mu},p_e)}{\ep^i},
\label{epexp}
\end{equation}
where the $A^{Y} _i$ are functions non-singular everywhere in phase-space.  Since the $A^{Y}_i$ are non-singular, they can be computed 
numerically in four dimensions.  The expressions for the $d\Gamma_{Y} / dx$ can be combined, and the poles in $\ep$ can be cancelled numerically.  
We must produce such an expansion for the following components.
\begin{enumerate}
  
\item The real radiation corrections involve decays with two additional particles radiated into 
    the final state.  The two relevant processes are $\mu \rightarrow e\nu\nu+\gamma\gamma$ and 
    $\mu \rightarrow e\nu\nu+e^+e^-$.  The first one begins at $1/\ep^2$, with the singularities 
    coming from the phase-space regions where the photon energies vanish, while the second is finite.  To 
    handle these corrections, we use the techniques presented in~\cite{secdecomp}.  We describe our 
    phase-space parameterizations and discuss the extraction of singularities in Subsection~\ref{subs.rr}.
  \item The real-virtual component includes the 1-loop virtual corrections to the process 
    $\mu \rightarrow e\nu\nu+\gamma$, and contributes beginning at $1/\ep^2$.  We use a hybrid approach combining both analytical and 
    numerical techniques to compute this component.  We first use integration-by-parts identities and recurrence 
    relations~\cite{tkachov} to reduce the corrections to a small set of master integrals.  The recurrence relations are 
    solved using the algorithm described in~\cite{laporta} and implemented in~\cite{air}.  We then solve 
    for the master integrals numerically by applying the techniques of~\cite{secdecomp} to their 
    Feynman parameter representation.  We discuss the details of this method, including how we handle 
    imaginary components of the loop integrals, in Subsection~\ref{subs.rv}.
  \item The virtual-virtual corrections contain the interference of two-loop virtual corrections to $\mu \rightarrow e\nu\nu$ 
    with tree-level diagrams.  These begin at $1/\ep^2$. 
    We deal with these completely numerically by applying 
    the techniques of~\cite{secdecomp} directly to their 
Feynman parameter representation.  This 
    numerical method of computing virtual corrections was 
pioneered in~\cite{Binoth:2000ps,Binoth:2003ak}.  We apply it here for 
    the first time in a fully realistic calculation, which includes tensor integrals and several mass scales.  We discuss 
    the details of this calculation in Subsection~\ref{2lvirt}.
  \item We must include the square of the NLO virtual corrections, which contribute at ${\cal O}(\alpha^2)$ and produce poles 
    beginning at $1/\ep^2$.  Since the computation of this component can be performed with standard techniques, 
    we do not discuss it further.
  \item We must include both fermion mass renormalization, and external wave-function renormalization.  We renormalize in the 
    on-shell scheme.  The renormalization is performed by multiplying 
    the LO and NLO results by the factor $Z_{2}^e \times Z_{2}^\mu$,
and by inserting the muon and electron  mass counterterms 
into the NLO 
diagrams. Therefore, the mass counter-term is needed 
through ${\cal O}(\alpha)$ only.  For a fermion of mass $m$,
the renormalization constants are 
\cite{broad,Melnikov:2000zc}
    \begin{eqnarray}
      Z_{2} &=&1+\sum_{n=1}
\left[\frac{\alpha}{\pi} 
\frac{\Gamma(1+e)  m^{-2\ep}}{(4\pi)^{-\ep}}\right]^n Z_{2}^{(n)}, \nonumber \\
      Z_{2}^{(1)}&=& -\frac{3}{4\ep}-\frac{1}{1-2\ep},~~~
      Z_{2}^{(2)}=\frac{9}{32\ep^2}+\frac{51}{64\ep}+\frac{433}{128}-\frac{3}{2}\zeta(3)+\pi^2 {\rm ln}(2)-\frac{13}{16}\pi^2,
\nonumber \\
\delta m & = & m_{\rm bare} - m = m(Z_m - 1) = 
\frac{\alpha}{\pi} 
\frac{\Gamma(1+e)  m^{1-2\ep}}{(4\pi)^{-\ep}}  Z_{2}^{(1)}.
    \end{eqnarray}
    %
We note that 
    contributions which arise from a closed fermion loop inserted into a 1-loop self-energy diagram, which appear at 
    ${\cal O}(\alpha^2)$, have been removed from these formulae; they are more naturally included in the contribution discussed in 
    the following item.
  \item Finally, we must include vacuum polarization corrections, in which a muon or electron loop is inserted into 
    a 1-loop diagram.  These include the insertion of a closed 
fermion loop into a 1-loop vertex diagram, and insertions into 
    external leg self-energy corrections, which contribute to 
the muon and electron wave function renormalization constants.
These corrections form 
    a finite subset.  They can be computed using dispersive techniques, as discussed in~\cite{vanRitbergen:1998hn,Davydychev:2000ee}, 
    where their contribution to the muon lifetime and electron energy spectrum are computed.   Since these corrections are 
    discussed in the literature, and can be computed with standard techniques, we do not discuss them further.
\end{enumerate}
After combining items $1-6$, both ultraviolet and infrared 
divergences cancel, leaving a finite result.
This completes the  brief summary of the terms which 
enter the ${\cal O}(\alpha^2)$ corrections; we now begin
the  technical discussion of their computation.

\subsection{Real radiation corrections}
\label{subs.rr}

\begin{figure}[htp]
\begin{center}
\begin{picture}(200,50)(0,0)
\SetColor{Blue}
\SetWidth{1.0}
\Line(0,35)(35,0)
\Line(35,0)(70,35)
\SetColor{Red}
\Vertex(35,0){3}
\SetColor{Black}
\Photon(5,30)(30,40){2}{4}
\Photon(15,20)(40,30){2}{4}
\put(0,40){$\mu$}
\put(70,40){$e$}

\SetColor{Blue}
\SetWidth{1.0}
\Line(110,35)(145,0)
\Line(145,0)(180,35)
\Line(137.5,25)(137.5,40)
\Line(137.5,25)(152.5,25)
\SetColor{Red}
\Vertex(145,0){3}
\SetColor{Black}
\Photon(125,20)(137.5,25){2}{2}
\put(110,40){$\mu$}
\put(180,40){$e$}

\end{picture}
\end{center}
\caption{\label{RRdiags} Sample diagrams which contribute  to $\mu \rightarrow e\nu\nu+\gamma\gamma$ (left) and 
$\mu \rightarrow e\nu\nu+e^+e^-$ (right)}
\end{figure}
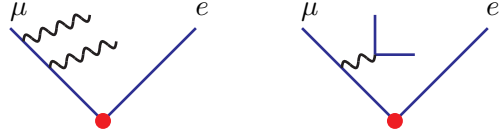

We first discuss the contribution of the real radiation processes $\mu \rightarrow e\nu\nu+\gamma\gamma$ and 
$\mu \rightarrow e\nu\nu+e^+e^-$.  A sample of the diagrams that contribute to these processes is shown in Fig.~\ref{RRdiags}.  
They produce a contribution to the differential decay rate of the form
\begin{equation}
\frac{d\Gamma_{RR}}{dx} =  \int dp_{\rm nt}^2 \int [d \Pi_{\mu \rightarrow ep_{\rm nt}+X}]  
 |{\cal M}^{\rho\sigma}_{\mu \rightarrow e+X}|^2 T^{\rho\sigma}_{\rm nt}.
\label{eq.rr.1}
\end{equation}

In order to perform the integration in Eq.~\ref{eq.rr.1}, 
we must discuss both our phase-space parameterizations and the singularity 
structure of the matrix elements.  Denoting the radiation momenta by $p_{\gamma 1}$, $p_{\gamma 2}$, $p_{e^{-}1}$, $p_{e^{+}2}$, 
we find the following denominators for each process:
\begin{itemize}
\item $\mu \rightarrow e\nu\nu+\gamma\gamma$: $d^{\gamma}_{\mu 1}=(p_{\mu}-p_{\gamma 1})-m_{\mu}^2$, $d^{\gamma}_{\mu 2}$, 
  $d^{\gamma}_{e1}=(p_{e}+p_{\gamma 1})-m_{e}^2$, $d^{\gamma}_{e2}$, $d^{\gamma}_{\mu 12}=(p_{\mu}-p_{\gamma 1}-p_{\gamma 2})-m_{\mu}^2$,
  $d^{\gamma}_{e12}=(p_{e}+p_{\gamma 1}+p_{\gamma 2})-m_{e}^2$;
\item $\mu \rightarrow e\nu\nu+e^+e^-$: $d^{e}_{12}=(p_{e^{-}1}+p_{e^{+}2})^2$, $d^{e}_{e2}=(p_{e}+p_{e^{+}2})^2$, 
  $d^{e}_{\mu 12}=(p_{\mu}-p_{e^{-}1}-p_{e^{+}2})^2-m_{\mu}^2$, $d^{e}_{\mu e2}=(p_{\mu}-p_{e}-p_{e^{+}2})^2-m_{\mu}^2$,
  $d^{e}_{e12}=(p_{e}+p_{e^{-}1}+p_{e^{+}2})^2-m_{e}^2$.
\end{itemize}
We first discuss our phase-space representation for the photon radiation process, which takes the form 
\begin{eqnarray}
 \int dp_{\rm nt}^2 & & \hspace{-0.5cm} \int [d \Pi_{\mu \rightarrow ep_{\rm nt}+X}]  =
 \frac{1}{(2\pi)^{3d-4}}  \int ds_{\rm nt} \int d^{d}p_{\rm nt} d^{d}p_{e} d^{d}p_{\gamma 1}d^{d}p_{\gamma 2 } \nonumber \\ 
  &\times&
  \delta(p_{\rm nt}^2-s_{\rm nt}) \delta(p_{e}^2 -m_{e}^2) \delta(p_{\gamma 1}^2) \delta(p_{\gamma 2}^2) \delta^{(d)}(p_{\mu}
  -p_{e}-p_{\gamma 1}-p_{\gamma 2}-p_{\rm nt}),
\end{eqnarray}
where the restriction of the electron energy fraction $x$ is understood in the rightmost equation.  It is convenient to evaluate 
this in the rest frame of the muon and to choose the $z$-axis 
along the electron direction.  In this frame, the momenta are 
\begin{eqnarray}
p_{\mu} &=& \left(m_{\mu},0,0,0\right), \;\;\; p_e = \left(E_e,0,0,\beta E_e\right), \nonumber \\
p_{\gamma 1} &=& \left(E_1, E_1 s_1, 0, E_1 c_1 \right), \;\;\;  p_{\gamma 2} = \left(E_2, E_2 s_2 c_{\phi}, 
  E_2 s_2 s_{\phi}, E_2 c_2 \right),
\end{eqnarray}
where $E_e$, $E_1$, and $E_2$ denote energies, $s_1$, $s_2$, $c_1$, and $c_2$ respectively denote sines and cosines of 
polar angles, and $s _{\phi}$, $c_{\phi}$ denote the sine and cosine of the azimuthal angle.  Following~\cite{secdecomp}, 
we map this to the unit hypercube, and obtain
\begin{eqnarray}
 N_{\gamma}& &  \hspace{-0.1cm}\int_{0}^{1} d\lambda_1 d\lambda_2 d\lambda_3 d\lambda_4 d\lambda_5 \, \kappa_{12}^{-2+2\ep} (1+\delta^2-x)^{4-4\ep}
  (1-\lambda_1)^{3-4\ep} [\beta x]^{1-2\ep} \nonumber \\ &&\times 
  \left[\lambda_2 (1-\lambda_2)\right]^{1-2\ep} \left[\lambda_3 \lambda_4 (1-\lambda_3)(1-\lambda_4)\right] ^{-\ep}
  \left[\lambda_5 (1-\lambda_5)\right]^{-1/2-\ep},
\label{gammaPS}
\end{eqnarray}
where
\begin{eqnarray}
 \delta &=& m_e / m_{\mu}, \;\;\; c_1 = 2\lambda_3-1, \;\;\; c_2 = 2\lambda_4-1, \;\;\; c_{\phi}=2\lambda_5-1, \nonumber \\ 
   E_e &=& x/2, \;\;\; E_1=\frac{\lambda_2 (1-\lambda_1) (1+\delta^2-x)}{2\kappa_1}, \;\;\; 
   E_2=\frac{\kappa_1 (1-\lambda_2) (1-\lambda_1) (1+\delta^2-x)}{2\kappa_{12}}, \nonumber \\
   N_{\gamma}&=& \frac{\Omega_{d-1}\Omega_{d-2}\Omega_{d-3}}{2^7 (2\pi)^{3d-4}}, \;\;\; \Omega_d = \frac{2\pi^{d/2}}{\Gamma(d/2)}, \;\;\;
   s_{\rm nt}=\lambda_1 (1+\delta^2-x), \;\;\; \kappa_1 =1-\frac{x}{2}(1-\beta c_1) \nonumber \\ 
   \kappa_{12}&=&\kappa_1- \frac{\kappa_1 x}{2}(1-\beta c_2)-\frac{\lambda_2 (1-\lambda_1)(1+\delta^2-x)(1-\vec{n}_1\cdot\vec{n}_2)}{2}, 
   \nonumber \\ \beta&=&\sqrt{1-4\delta^2/x^2},\;\;\; \vec{n}_1\cdot\vec{n}_2 = c_1 c_2 + s_1 s_2 c_{\phi}.
\end{eqnarray}
We have removed the integration over the energy fraction $x$, and have set the scale $m_{\mu}=1$; it can be restored with dimensional 
analysis.  The matrix element denominators become
\begin{eqnarray}
  d^{\gamma}_{\mu 1}&=&\frac{-\lambda_2 (1-\lambda_1)(1+\delta^2-x)}{\kappa_1}, \;\;\; 
  d^{\gamma}_{\mu 2}=\frac{-\kappa_1 (1-\lambda_2) (1-\lambda_1)(1+\delta^2-x)}{\kappa_{12}}, \nonumber \\ 
  d^{\gamma}_{e1}&=&\frac{x \lambda_2 (1-\lambda_1)(1+\delta^2-x)(1-\beta c_1)}{2\kappa_1}, \;\;\; 
  d^{\gamma}_{e2}=\frac{x \kappa_1 (1-\lambda_2) (1-\lambda_1)(1+\delta^2-x)(1-\beta c_2)}{2\kappa_{12}}, \nonumber \\ 
  d^{\gamma}_{\mu 12}&=& -(1-\lambda_1)(1+\delta^2-x)\left\{1+\frac{x\lambda_2 (1-\beta c_1)}{2\kappa_1} 
   +\frac{x\kappa_1 (1-\lambda_2) (1-\beta c_2)}{2\kappa_{12}} \right\}, \nonumber \\ 
  d^{\gamma}_{e12}&=& (1-\lambda_1)(1+\delta^2-x)\left\{-1 +\frac{\lambda_2}{\kappa_1}+\frac{\kappa_1 (1-\lambda_2)}{\kappa_{12}} \right\}.
\end{eqnarray}
We note that all the divergences are produced by the overall multiplicative factors of $(1-\lambda_1)$, $\lambda_2$, and 
$(1-\lambda_2)$; the bracketed terms in $d^{\gamma}_{\mu 12}$ and $d^{\gamma}_{e12}$ are finite for values of $x$ away from its 
boundaries.  In the language of~\cite{secdecomp}, all singularities are {\it factorizable}; when the denominators are combined with the 
phase-space in Eq.~\ref{gammaPS}, the form of Eq.~\ref{epexp} can be produced by expanding in plus distributions:
\begin{eqnarray}
\lambda^{-1+e}&=& \frac{1}{\ep}\delta(\lambda) +\sum_{n=0} \left[\frac{{\rm ln}^n(\lambda)}{\lambda}\right]_{+}\frac{\ep^n}{n!}, \nonumber \\ 
&& \int_{0}^{1} d\lambda \, f(\lambda)\left[\frac{{\rm ln}^n(\lambda)}{\lambda}\right]_+ = \int_{0}^{1} d\lambda \,\frac{f(\lambda)-f(0)}{\lambda}\,
{\rm ln}^{n}(\lambda).
\label{plusexp}
\end{eqnarray}

We must now discuss our phase-space representation for the process $\mu \rightarrow e\nu\nu+e^+e^-$, which takes the form 
\begin{eqnarray}
 \int dp_{\rm nt}^2 & & \hspace{-0.5cm} \int [d \Pi_{\mu \rightarrow ep_{\rm nt}+X}]  =
 \frac{1}{(2\pi)^{3d-4}}  \int ds_{\rm nt} \int d^{d}p_{\rm nt} d^{d}p_{e} d^{d}p_{e^- 1}d^{d}p_{e^+ 2 } \,\delta(p_{\rm nt}^2-s_{\rm nt}) \nonumber \\ 
  &\times&
  \delta(p_{e}^2 -m_{e}^2) \delta(p_{e^- 1}^2-m_e^2) \delta(p_{e^+ 2}^2-m_e^2) \delta^{(d)}(p_{\mu}
  -p_{e}-p_{e^- 1}-p_{e^+ 2}-p_{\rm nt}).
\end{eqnarray}
It is convenient to view this decay as occurring iteratively; first, the muon decays into an electron and a massive ``particle'' 
with momentum $p_{\rm nt}+p_{e^- 1}+p_{e^+ 2}$.  This massive particle then decays into $p_{e^- 1}$ and another massive 
particle with momentum $p_{\rm nt}+p_{e^+ 2}$, which finally decays into $p_{e^+ 2}$ and the neutrino pair.  This motivates the 
following decomposition of the phase-space:
\begin{equation}
\frac{1}{(2\pi)^{3d-4}} \int ds_{\rm nt} ds_{\rm nt 12} ds_{\rm nt 2} \, I_1 I_2 I_3,
\end{equation}
where 
\begin{eqnarray}
  I_1 &=& \int d^{d}p_{e} d^{d}p_{\rm nt 12} \,\delta(p_{e}^2 -m_{e}^2) \,\delta(p_{\rm nt 12}^2 -s_{\rm nt 12}) \,\delta^{(d)}(p_{\mu}-p_e -p_{\rm nt 12}), 
    \nonumber \\
  I_2 &=& \int d^{d}p_{e^- 1} d^{d}p_{\rm nt 2} \,\delta(p_{e^- 1}^2 -m_{e}^2) \,\delta(p_{\rm nt 2}^2 -s_{\rm nt 2}) \,\delta^{(d)}
    (p_{\rm nt 12}-p_{e^- 1} -p_{\rm nt 2}), \nonumber \\
  I_3 &=& \int d^{d}p_{e^+ 2} d^{d}p_{\rm nt} \,\delta(p_{e^+ 2}^2 -m_{e}^2) \,\delta(p_{\rm nt}^2 -s_{\rm nt})\,
    \delta^{(d)}(p_{\rm nt 2}-p_{e^+ 2} -p_{\rm nt}).
\end{eqnarray}
We evaluate $I_1$, $I_2$, and $I_3$ in the rest frame of the massive ``particle'' that defines each phase-space.  Doing so yields the 
following expression:
\begin{eqnarray}
  N_{ee}& &  \hspace{-0.1cm}\int_{0}^{1} d\lambda_1 d\lambda_2 d\lambda_3 d\lambda_4 d\lambda_5 \, \frac{\Delta_{34}\left(\sqrt{1+\delta^2-x}-2\delta\right)^2}
    {\sqrt{s_{\rm nt 12}s_{\rm nt 2}}} \left[\beta^3 x E_2 E_3\right]^{1-2\ep} \nonumber \\ 
    && \times  \left[\lambda_3 \lambda_4 (1-\lambda_3)(1-\lambda_4)\right] ^{-\ep}
  \left[\lambda_5 (1-\lambda_5)\right]^{-1/2-\ep},
\label{eq.rr.2}
\end{eqnarray}
where 
\begin{eqnarray}
  N_{ee}&=& \frac{\Omega_{d-1}\Omega_{d-2}\Omega_{d-3}}{2^{5+4\ep} (2\pi)^{3d-4}}, \;\;\; 
    E_2 = \frac{s_{\rm nt 12}+\delta^2-s_{\rm nt 2}}{2\sqrt{s_{\rm nt 12}}}, \;\;\; 
    E_3 = \frac{s_{\rm nt 2}+\delta^2-s_{\rm nt}}{2\sqrt{s_{\rm nt 2}}}, \;\;\; \nonumber \\ 
  s_{\rm nt 12}&=& 1+\delta^2-x,\;\;\; s_{\rm nt}=\lambda_1\left(\sqrt{1+\delta^2-x}-2\delta\right)^2,\;\;\; 
    s_{\rm nt 2}=\Delta_{34} \lambda_2 +L_{34}, \nonumber \\
  \Delta_{34}&=& \left(\sqrt{s_{\rm nt 12}}-\delta\right)^2-\left(\sqrt{s_{\rm nt}}+\delta\right)^2,\;\;\;
    L_{34}=\left(\sqrt{s_{\rm nt 12}}-\delta\right)^2, \nonumber \\
  c_2 &=& 2\lambda_3-1, \;\;\; c_3 = 2\lambda_4-1, \;\;\; c_{\phi}=2\lambda_5-1.
\end{eqnarray}
$c_2$ is the polar angle of $p_{e^- 1}$ within the frame of $I_2$, while $c_3$ and $c_{\phi}$ are respectively the 
polar and azimuthal angles of $p_{e^+ 2}$ within $I_3$.  To obtain the invariant masses that appear in the matrix 
elements, we must Lorentz transform the vectors $p_{\mu}$, $p_{e}$, $p_{e^- 1}$, and $p_{e^+ 2}$ between the frames 
defined by the three phase-spaces.  We note that all of the denominators that appear in the 
matrix elements are regulated by $\delta$, and therefore the process $\mu \rightarrow e\nu\nu+e^+e^-$ is finite.  It is sufficient 
to set $\epsilon = 0$ in Eq.~\ref{eq.rr.2} and to perform the numerical integration of the corresponding matrix element in four 
dimensions.

\subsection{Virtual corrections to single photon emission}
\label{subs.rv}

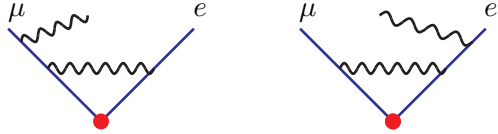
\begin{figure}[htp]
\begin{center}
\begin{picture}(200,50)(0,0)
\SetColor{Blue}
\SetWidth{1.0}
\Line(0,35)(35,0)
\Line(35,0)(70,35)
\SetColor{Red}
\Vertex(35,0){3}
\SetColor{Black}
\Photon(5,30)(30,40){2}{4}
\Photon(15,20)(55,20){2}{6}
\put(0,40){$\mu$}
\put(70,40){$e$}

\SetColor{Blue}
\SetWidth{1.0}
\Line(110,35)(145,0)
\Line(145,0)(180,35)
\SetColor{Red}
\Vertex(145,0){3}
\SetColor{Black}
\Photon(125,20)(165,20){2}{6}
\Photon(175,30)(140,40){2}{4}
\SetColor{Red}
\Vertex(145,0){3}
\put(110,40){$\mu$}
\put(180,40){$e$}

\end{picture}
\end{center}
\caption{\label{RVdiags} Sample diagrams which contribute  to $\mu \rightarrow e\nu\nu+\gamma$ }
\end{figure}

Here we discuss the one-loop virtual correction to the 
process $\mu \to e \nu \nu + \gamma$; some  diagrams that must be 
considered are shown in Fig.~\ref{RVdiags}.
This computation is conveniently performed 
 by a combination of analytical and 
numerical methods. First, we express the integrals over the loop momenta
through master integrals using the reduction algorithm of \cite{laporta} 
implemented in \cite{air}. To give the list of master integrals, we introduce 
the following notation: 
\begin{eqnarray}
&&{\rm DP}_1(a_1,a_2,a_3,a_4) = 
\int \frac{{\rm d}^d k}{(2\pi)^d} 
\frac{1}{(k^2+2p_\mu k)^{a_1} ((k+p_\mu - p_\gamma)^2 - m_\mu^2)^{a_2}
(k^2 + 2p_e k)^{a_3} k^{2a_4}},
\nonumber \\
&& 
\label{eq.rv.1}
\\
&&{\rm DP}_2(a_1,a_2,a_3,a_4) = 
\int \frac{{\rm d}^d k}{(2\pi)^d} 
\frac{1}{(k^2+2p_\mu k)^{a_1} ((k+p_e + p_\gamma)^2 - m_e^2)^{a_2}
(k^2 + 2p_e k)^{a_3} k^{2a_4}}.
\nonumber
\end{eqnarray}
We find that all the Feynman integrals needed for our purposes are expressed 
through fifteen master integrals that include the four-point functions 
${\rm DP}_1(1,1,1,1)$ and ${\rm DP}_2(1,1,1,1)$, several three-point functions
such as ${\rm DP}_1(0,1,1,1),{\rm DP}_1(1,1,1,0),{\rm DP}_2(1,1,0,1)$
and a number of two-point functions and tadpoles. These Feynman integrals
depend on the energy of the external photon, $\omega_\gamma$; when $\omega_\gamma \to 0$, some 
of the master integrals develop infrared singularities. 
The extraction of singularities is
performed following~\cite{secdecomp}; we 
Feynman-parameterize the master integrals, insert these expressions  
into our phase-space parameterization, and disentangle singularities in 
both the Feynman parameters and $\omega_\gamma$.

We write the real-virtual component of the NNLO corrections as 
\begin{equation}
\frac{d\Gamma_{RV}}{dx} =  \int dp_{\rm nt}^2 
\int [d \Pi_{\mu \rightarrow ep_{\rm nt}+\gamma}]  
 |{\cal M}^{\rho\sigma}_{\mu \rightarrow e+\gamma}|^2 T^{\rho\sigma}_{\rm nt}.
\label{eq.rv.1.5}
\end{equation}
It is quite easy to construct a phase-space parameterization 
convenient for the extraction of singularities; it is very similar 
to the double real emission case discussed in the 
previous Subsection. We find
\be
\int dp_{\rm nt}^2 
\int [d \Pi_{\mu \rightarrow ep_{\rm nt}\gamma}] 
= \frac{\Omega_{d-1} \Omega_{d-2}}{2^{3+2\epsilon}(2\pi)^{2d-3}}
\int \limits_{0}^{1} 
{\rm d}\lambda_1  
{\rm d}\lambda_2 \lambda_1^{1-2 \ep} \lambda_2^{-\ep} (1-\lambda_2)^{-\ep}
{\cal F}_{\rm rv}(x,\lambda_1,\lambda_2),
\label{eq.rv.1.6}
\ee
where 
\be
{\cal F}_{\rm rv} = \frac{(E_{\rm max}^2 z (1-z) 
\beta)^{1-2\ep}}{(1-E_e(1-\beta \cos \theta))^{2-2\ep}},
\ee
$E_{\rm max} = (1+\delta^2)/2$, $E_e = x/2$, $ z = E_e/E_{\rm max}$,
$p_{\rm nt}^2 = 2E_{\rm max} (1-z)(1-\lambda_1)$, and  $\cos \theta = -1+2\lambda_2$. In terms of these variables, 
the scalar products of the four-momenta  $s_{ab} = 2 p_a \cdot p_b$
read 
\be
s_{\mu e} =2E_e,~~~~~s_{\mu \gamma} = 2 \omega_\gamma = 
\frac{2 E_{\rm max}(1-z) \lambda_1}{1-E_e(1-\beta  \cos \theta)},~~~~
s_{e \gamma} = 2E_e \omega_\gamma (1-\beta \cos \theta). 
\label{eq.rv.1.7}
\ee
 From 
Eqs.~\ref{eq.rv.1.6} and~\ref{eq.rv.1.7} we see that potential singularities 
associated 
with the soft photon emission $\omega_\gamma \to 0$ are factorized 
both in the phase-space and in the scalar products $s_{ab}$; therefore, 
their extraction proceeds along the lines described in \cite{secdecomp}.

An additional complication related to the real-virtual corrections is that
some of the master integrals develop imaginary parts that, when the 
integral is Feynman-parameterized, appear as singularities 
in the integration region. This feature is very inconvenient since 
it makes it impossible to numerically integrate even 
otherwise finite expressions. We explain how we deal with 
this problem by considering the master integral 
${\rm DP}_2(1,1,1,1)$ of Eq.~\ref{eq.rv.1}.

We introduce a Feynman parameterization for the integral ${\rm DP}_2(1,1,1,1)$, and write it as
\be
{\rm DP}_2(1,1,1,1) = \frac{i \Gamma(2+\epsilon)}{(4\pi)^{d/2}}
\int \limits_{0}^{1}  {\rm d} 
\lambda_3 \prod \limits_{i=1..3}^{} {\rm d} x_i
\delta(1 - \sum \limits_{i=1}^{3}x_i) \frac {x_2}{\phi^{2+\epsilon}}, 
\ee
where $\phi = -s_{e\gamma} \lambda_3 x_2 + x_1^2 + (m_e^2 + s_{e \gamma} \lambda_3 )x_2^2 
+x_1 x_2 (s_{\mu e} + \lambda_3 s_{\mu \gamma} )$.
Changing variables to $x_{2} = \lambda_1 \lambda_2$
and $x_1 = \lambda_1 (1-\lambda_2)$, we find 
\be
{\rm DP}_2(1,1,1,1) = \frac{i \Gamma(2+\epsilon)}{(4\pi)^{d/2}}
\int \limits_{0}^{1}  {\rm d} 
\lambda_3 {\rm d}\lambda_1 {\rm d} \lambda_2
\frac {\lambda_1^{-\epsilon} \lambda_2 }
{\left ( \lambda_1 \Delta - s_{e\gamma} \lambda_3 \lambda_2 \right )^{2+\epsilon}}, 
\label{eq.rv.2}
\ee
where $\Delta = (1-\lambda_2)^2 + (m_e^2 + s_{e \gamma}\lambda_3)  \lambda_2^2 
+ \lambda_2 (1-\lambda_2) (s_{\mu e } + \lambda_3 s_{\mu \gamma}) $. 
The denominator 
in Eq.~\ref{eq.rv.2} can become zero in 
the integration region, which makes accurate numerical 
evaluation impossible even for non-exceptional values of the photon energy.
To circumvent this problem, we rewrite Eq.~\ref{eq.rv.2} in the following 
way. First, we integrate over $\lambda_1$, producing a hypergeometric function
\be
{\rm DP}_2(1,1,1,1) = \frac{i \Gamma(2+\epsilon)}{(4\pi)^{d/2}}
\int \limits_{0}^{1}  {\rm d} 
\lambda_3  {\rm d} \lambda_2~\lambda_2 (-s_{e\gamma} \lambda_3 \lambda_2)^{-2-\epsilon} 
\frac{\Gamma(1-\epsilon)}{\Gamma(2-\epsilon)}
F_{21}\left(2+\epsilon,1-\epsilon;2-\epsilon,
\frac{\Delta}{s_{e\gamma} \lambda_3 \lambda_2} \right ).
\ee
We now use an identity that allows us to rewrite $F_{21}(a,b,c,z)$ through 
$F_{21}(...,1/z)$. We obtain
\begin{eqnarray}
&& {\rm DP}_2(1,1,1,1) = \frac{i \Gamma(2+\epsilon)}{(4\pi)^{d/2}}
\int \limits_{0}^{1}  {\rm d} 
\lambda_3  {\rm d} \lambda_2~\lambda_2 (-s_{e\gamma} \lambda_3 \lambda_2)^{-2-\epsilon} 
\left \{
\frac{-1}{1+2\epsilon} \left (\frac{\Delta}{-s_{e\gamma} \lambda_3 \lambda_2} 
\right )^{-2-\epsilon}
\right. 
\nonumber \\
&& \left. \times F_{21} \left (2+\epsilon,1+2\epsilon,2+2\epsilon,
\frac{s_{e\gamma}\lambda_3 \lambda_2}{\Delta} \right )
+ \frac{\Gamma(1-\epsilon) \Gamma(1+2\epsilon)}{\Gamma(2+\epsilon)}
\left (\frac{\Delta}{-s_{e\gamma} \lambda_3 \lambda_2} \right )^{-1+\epsilon}
\right \}.
\label{eq.rv.4}
\end{eqnarray}
For the hypergeometric function that appears in Eq.~\ref{eq.rv.4} 
we introduce a standard integral representation and arrive at
\begin{eqnarray}
{\rm DP}_2(1,1,1,1) &=& \frac{-i \Gamma(2+\epsilon)}{(4\pi)^{d/2}}
\int \limits_{0}^{1}  \prod \limits_{i=1}^{3} {\rm d}\lambda_i
\left \{ 
 \frac{\lambda_2 \lambda_1^{2\epsilon}}{(\Delta - s_{e \gamma} \lambda_2 \lambda_1 \lambda_3)^{2 + \epsilon}} \right. \nonumber \\ 
&+& \left. \frac{\Gamma(1-\epsilon)\Gamma(1+2\epsilon)}{\Gamma(2+\epsilon)}
\frac{(-s_{e\gamma} \lambda_3 \lambda_2)^{-2\epsilon}}{s_{e\gamma} \lambda_3 \Delta^{1-\epsilon}}
\right \}.
\label{eq.rv.5}
\end{eqnarray}
It is easy to see that the denominator of the first term on the right-hand side of
Eq.~\ref{eq.rv.5} does not vanish inside the integration region; 
the imaginary part appears only from the second term on the right-hand side of 
Eq.~\ref{eq.rv.5}, which is  $\propto (-1)^{-2\epsilon}$. Hence, Eq.~\ref{eq.rv.5} 
can be used for numerical integration after disentangling singularities 
in $\omega_{\gamma}$ and the Feynman parameters.

\subsection{Two-loop virtual corrections}
\label{2lvirt}

We compute the NNLO two-loop diagrams  numerically. A basic ingredient 
of this approach is  the method of sector decomposition. In the past, 
this technique has been applied to scalar loop-integrals. In this paper, 
we extend the approach to compute a full two-loop amplitude; to the 
best of our knowledge, this is done here for the first time.. 
The tensor integrals that emerge in the two-loop amplitude 
can be expressed in terms of scalar integrals using, for example,
the procedure in~\cite{pentabox,tarasov}. In principle, the latter could  
be computed with a brute force application of the algorithm 
in~\cite{Binoth:2000ps,Binoth:2003ak}. 
However, we have found that it is more efficient 
to adopt a slight modification of that approach. Specifically, 
since the number of Feynman diagrams we deal with is not large, 
we derive a Feynman integral representation for each of the diagrams. 
Such  representations are not unique; the ones we derive simplify the evaluation of tensor integrals. First, we introduce a  Feynman 
parameterization for the propagators of one of the loop integrals.  We then 
integrate out the corresponding loop-momentum, and insert the result into 
the second loop. We carry out the remaining loop integration with
a new set of Feynman parameters, using the approach of ~\cite{Binoth:2000ps,Binoth:2003ak}.
 
\begin{figure}[htp]
\begin{center}
\begin{picture}(80, 50)(0, 0)
\SetColor{Blue}
\SetWidth{1.0}
\Line(0,35)(35,0)
\Line(35,0)(70,35)
\SetColor{Red}
\Vertex(35,0){3}
\SetColor{Black}
\Photon(5, 30)(65,30){2}{6}
\Photon(15, 20)(55,20){2}{4}
\put(0,40){$\mu$}
\put(70,40){$e$}
\end{picture}
\hspace{0.5cm}
\begin{picture}(80, 50)(0, 0)
\SetColor{Blue}
\SetWidth{1.0}
\Line(0,35)(35,0)
\Line(35,0)(70,35)
\SetColor{Red}
\Vertex(35,0){3}
\SetColor{Black}
\Photon(5, 30)(55,20){2}{5}
\Photon(15, 20)(65,30){2}{5}
\put(0,40){$\mu$}
\put(70,40){$e$}
\end{picture}
\hspace{0.5cm}
\begin{picture}(80, 50)(0, 0)
\SetColor{Blue}
\SetWidth{1.0}
\Line(0,35)(35,0)
\Line(35,0)(70,35)
\SetColor{Red}
\Vertex(35,0){3}
\SetColor{Black}
\Photon(10, 25)(60,25){2}{5}
\PhotonArc(10, 25)(8,130,310){1}{5}
\put(0,40){$\mu$}
\put(70,40){$e$}
\end{picture}
\end{center}
\caption{\label{effVV} A sample of two-loop diagrams which contribute to $\mu \rightarrow e\nu\nu$.}
\end{figure}
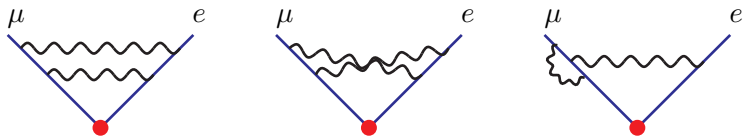

As an example, we  derive the parameterization for tensor integrals in the cross-triangle topology, which is the 
second diagram in Fig.~\ref{effVV}.. We consider the 
integral 
 \begin{equation}
 \label{eq:xtriangle}
{\cal X} = \int \frac{d^dk_1}{i \pi^{d/2}} \frac{d^dk_2}{i \pi^{d/2}}
\frac{\left\{ k_1 \right\}_m  \left\{ k_2 \right\}_n}
{A_1 A_2 A_3 A_4 A_5 A_6},
 \end{equation}
 where 
 \begin{eqnarray}
&&  A_1 = k_1^2,~~~~A_2 = k_2^2,~~~A_3 = k_1^2 + 2 k_1\cdot p_\mu, 
~~~~A_4 = k_2^2+2 k_2 \cdot p_e,
\nonumber \\
&& A_5 = \left( k_1+k_2\right)^2 + 2  \left( k_1+k_2\right) \cdot p_\mu, 
~~~~A_6 = \left( k_1+k_2\right)^2 + 2  \left( k_1+k_2\right) \cdot p_e,
\end{eqnarray}
 and $p_\mu^2=m_\mu^2 = 1, p_e^2=m_e^2$. We denote a tensor of rank $m$ in the numerator with  $\left\{ k \right\}_m \equiv
 k^{\mu_1 \ldots \mu_m}$.
We first introduce Feynman parameters for the propagators in the $k_2$ loop. We write
 \begin{equation}
 \frac{1}{A_2 A_4 A_5 A_6} = \Gamma(4)\int_0^1 \frac{ d\lambda_1 d\lambda_2 d\lambda_3 \lambda_3 (1-\lambda_3)}
 {\left[ (k_2+q)^2 - \lambda_3 (1-\lambda_3) {\cal C_\alpha}\right]^4},
 \end{equation}
 with 
 \begin{eqnarray}
q &=& \lambda_3 k_1 + \eta, \nonumber \\
\eta &=& \lambda_3 \left[  \lambda_1p_\mu + (1-\lambda_1) p_e \right] + (1-\lambda_3) \lambda_2 p_e, \nonumber \\
 {\cal C_\alpha} &=& k_1^2 + 2 k_1\cdot \rho - \frac{ \eta^2 }{ \lambda_3 (1-\lambda_3)}, \nonumber \\
\rho  &=& \left[ \lambda_1 p_\mu +(1-\lambda_1-\lambda_2)p_e \right].
\end{eqnarray}
We then shift the momentum $k_2$, 
 \begin{equation}
 \label{eq:shift1}
 k_2 =K - q;
\end{equation}
the shift  yields a sum of tensors in $K$ with 
ranks $i \leq n$:
\begin{equation}
\left\{ k_2 \right\}_n  \to \sum_{i \leq n} { c}_i \left\{ K \right\}_i. 
\end{equation}
It is now straightforward to integrate out the loop-momentum $K$, using 
\begin{equation}
 \int \frac{d^d K}{i \pi^{d/2}} 
 \frac{ \left\{ K\right\}_n}{(K^2 + \Delta)^\alpha} = 
 \left( -1\right)^{d/2} \frac{\Gamma \left( \alpha- \frac{d+n}{2}\right)}{ 2^n \Gamma \left( \alpha \right) } \Delta^{\frac{d+n}{2}-\alpha} {\cal T}_n.
 \end{equation} 
 ${\cal T}_n=0$ for odd $n$, and ${\cal T}_0 =1, {\cal T}_2 =g^{\mu_1 \mu_2}, 
 {\cal T}_4 =g^{\mu_1 \mu_2}g^{\mu_3 \mu_4}+g^{\mu_1 \mu_3}g^{\mu_2 \mu_4}+g^{\mu_1 \mu_4}g^{\mu_2 \mu_3}, \mbox{etc} $. 
In order to perform the $k_1$ integration we introduce a new set of Feynman 
parameters $\lambda_4,\lambda_5$ and shift the momentum $k_1$. 
The shift yields a new set of 
terms which, after the integration, become 
\begin{equation}
\label{eq:tensor}
 {\cal X}_{ij} = \Gamma \left(2+2\epsilon -\frac{i+j}{2} \right) {\cal T}_i
{\cal T}_j
\int_0^1  \left( \prod_{k=1}^5 d\lambda_k\right)
\frac{(1-\lambda_5) \lambda_5^{1+\epsilon-i/2} \left[ \lambda_3 (1-\lambda_3)\right]^{1+\epsilon-j/2}}
{{\cal F}^{2+2\epsilon - \frac{i+j}{2}}}, 
\end{equation}
with 
\begin{equation}
{\cal F} = \lambda_5 \eta^2 + \lambda_3 (1-\lambda_3)\left[ 
p_\mu \lambda_4 (1-\lambda_5) + \lambda_5 \rho
\right]^2 
\end{equation}
The tensor integral is now written as 
 \begin{equation}
\label{eq:alltensor}
 {\cal X} = \sum_{i,j} f_{ij}\left(\lambda_1,\ldots,\lambda_5;p_\mu,p_e \right) 
{\cal X}_{ij},
 \end{equation}
where the terms $f_{ij}$ are polynomials in the Feynman parameters; they 
are produced from the shifts of the loop momenta. 
The above Feynman representation is ideal for integrating over Feynman 
parameters after the sector decomposition in~\cite{Binoth:2000ps,Binoth:2003ak}
is applied.   Explicit expressions for the rather lengthy polynomials 
$f_{ij}$ in  Eq.~\ref{eq:alltensor} are not required in order to write 
down  a Laurent expansion in $\epsilon$; these functions are only 
used in the Fortran code where coefficients of the $\epsilon$ expansion 
are evaluated. We emphasize that our parameterization is very convenient 
since it allows us to treat tensor integrals 
on the same footing as scalar integrals. 

\begin{figure}[htb]
\begin{center}
\begin{picture}(80, 50)(0, 0)
\SetColor{Blue}
\SetWidth{1.0}
\Line(0,35)(35,0)
\Line(35,0)(70,35)
\SetColor{Red}
\Vertex(35,0){3}
\SetColor{Black}
\Photon(5, 30)(65,30){2}{6}
\PhotonArc(15, 20)(8,130,310){1}{5}
\put(0,40){$\mu$}
\put(70,40){$e$}
\end{picture}
\end{center}
\caption{\label{fig.vv.2} A two-loop diagram with a self-energy insertion.}
\end{figure}
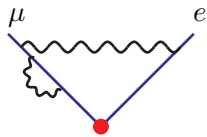

A technical complication arises when we consider two-loop diagrams with a self-energy insertion 
on either the muon or electron line, as in Fig.~\ref{fig.vv.2}.  Quadratic singularities of the 
form $\lambda^{-2-e}$ are produced, where $\lambda$ denotes one of the Feynman parameters.  We cannot use the 
expansion of Eq.~\ref{plusexp} to extract these singularities as a Laurent expansion in $\ep$.  
This occurs because one of the propagators in 
such diagrams appears squared. We solve this problem with the following 
procedure. As before, we start by  performing one of the loop integrations; 
here it is easy to integrate out the self-energy. 
Let $p$ be the momentum  entering the self-energy loop 
and $m$ the mass in the propagator with momentum $p$. The result of the 
integration is 
\begin{equation}
S = \frac{(-1)^{-\ep}\Gamma(1+\ep)}{\ep}\int_0^1 d\lambda_1 \frac{f(\lambda_1) \lambda_1^{-\epsilon} (1-\lambda_1)^{-\epsilon}}
{\left( p^2-m^2-\frac{\lambda_1 m^2}{1-\lambda_1} \right)^{\epsilon}},
\end{equation}
where $f$ is the polynomial from tensor reduction. We then insert this result 
into the second loop integration and  
obtain  the following structure in the denominator
\begin{equation}
{\Lambda} = \frac{1}{\left( p^2-m^2\right)^2 \left( p^2-m^2-\frac{\lambda_1 m^2}{1-\lambda_1} \right)^{\epsilon}}.
\end{equation}
A direct Feynman parameterization of $\Lambda$ and sector decomposition 
produces quadratic singularities. We avoid this by writing
\begin{equation}
\label{eq:selfentrick}
\Lambda =  \frac{\left(\frac{\lambda_1 m^2}{1-\lambda_1}\right)^{-\epsilon}}{(p^2-m^2)^2} 
- \epsilon \int_0^1 d\lambda_2 \frac{\lambda_2^{-1-\epsilon}}{
(p^2-m^2) \left( p^2-m^2-\frac{\lambda_1 m^2}{(1-\lambda_1)\lambda_2}\right)^{1+\epsilon}
}
\end{equation}
The first term in Eq.~\ref{eq:selfentrick} leads to a straightforward 
one-loop integration, since all propagators are raised to integer powers. 
In the second term, the offending propagator is not raised to a quadratic 
power anymore. We employ a parameterization of this term following a similar
procedure as in the cross-triangle topology discussed above.    

To summarize, we derive representations for  the two-loop diagrams in the 
process $\mu \to e \nu \nu$ which (i) are amenable to sector decomposition, (ii) treat tensor and scalar integrals on the same footing, 
and (iii) are 
free of quadratic singularities. We then produce an $\epsilon$-expansion of 
the diagrams using Eq.~\ref{plusexp}, and finally we evaluate the 
coefficients of the expansion numerically.

\section{Results}
\label{sect.res}

In this Section we  
give numerical results for the ${\cal O}(\alpha^2)$ 
corrections to the electron energy spectrum in unpolarized muon decay.  
We present our results in the form of a relative correction, 
$\delta^{(2)}=(\alpha / \pi)^2 f^{(2)}(x)/f^{(0)}(x)$, 
where $f^{(0)}$ and $f^{(2)}$ are 
respectively the LO and NNLO coefficient functions defined 
in Eq.~\ref{decaydef}.  This form allows us to study the 
magnitude of the corrections with respect to the relative experimental 
precision.  We employ the numerical values $m_{\mu} = 105.658357$ MeV 
and $m_{e} = 0.510998902$ MeV and the on-shell value
of the QED coupling constant, 
$\alpha=1/137.0359895$. For numerical estimates we consider 
electron energies in the range $0.3 \leq x \leq 0.95$, which
matches the acceptance of the TWIST experiment~\cite{TWIST}.  We have 
checked that integrating our result over $x$ reproduces the correction 
to the total decay rate found in~\cite{QED2loop}, within numerical integration errors.


As mentioned in the Introduction, 
the decay spectrum contains logarithms of the 
form ${\rm ln}(m_{\mu}/m_e)$, indicating that the electron energy 
is not physically observable as $m_e \rightarrow 0$.  The NNLO coefficient function can be expanded as a 
series in this logarithm:
\begin{equation}
f^{(2)}(x) = {\rm ln}^2(m_{\mu} /m_e)\,  f^{(2)}_2(x) 
+{\rm ln}(m_{\mu} /m_e) \,f^{(2)}_1(x) 
+f^{(2)}_0(x).
\label{logexp}
\end{equation}
The $f^{(2)}_2(x)$ and $f^{(2)}_1(x)$ terms have been calculated previously in~\cite{Arbuzov:2002pp,Arbuzov:2002cn}.  The new 
result of this paper is the  term $f^{(2)}_0(x)$.  
The uncertainty associated with the impact of $f^{(2)}_0(x)$ on the electron energy spectrum 
was previously estimated as $\approx 10^{-4}$, and its computation 
is necessary to match 
the precision expected in the TWIST experiment.

\begin{figure}[ht]
\centerline{
\psfig{figure=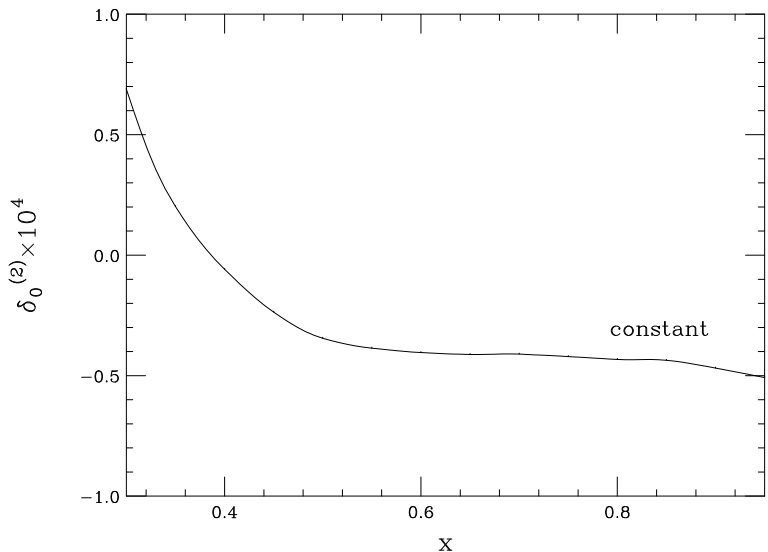,width=4.0in,angle=90}
}
   \caption{\label{constantterm} The ratio of the constant NNLO coefficient relative to the tree result, 
     $\delta^{(2)}_0=(\alpha / \pi)^2 f^{(2)}_0(x)/f^{(0)}(x)$, versus the electron energy fraction $x$.  The $y$-axis has been
     scaled by $10^4$.
}
\end{figure}

The magnitude of  $f^{(2)}_0(x)$  relative to the 
tree-level result as a function of the electron energy fraction $x$ 
is presented in Fig.~\ref{constantterm}.  To derive $f^{(2)}_0$, we calculate $f^{(2)}$ using our numerical 
program, and subtract from it the logarithmically enhanced terms given in~\cite{Arbuzov:2002pp,Arbuzov:2002cn}.  
We see that for a large range of electron energies, 
the absolute value of  $f^{(2)}_0(x)$ is bounded by $0.5 \times 10^{-4}$, 
a value somewhat smaller  than the theoretical expectations 
\cite{Arbuzov:2002pp,Arbuzov:2002cn}. To estimate the remaining 
theoretical uncertainty on the electron spectrum, we note that 
${\cal O}(\alpha^3 \ln^3(m_\mu/m_e))$ corrections to the spectrum 
have been computed in \cite{arbuzovh}; for moderate values of $x$, 
the corrections are in the range of ${\rm few} \times 10^{-6}$. 
The pattern of logarithmic corrections at ${\cal O}(\alpha^2)$ indicates that the
${\cal O}(\alpha^3 \ln^2(m_\mu/m_e))$ terms might have a similar size. The hadronic 
correction to the electron energy spectrum considered 
in \cite{Davydychev:2000ee} is even smaller. 
Similarly, finite $W$-mass effects are known, and 
influence the electron energy spectrum at the level of $\sim 10^{-6}$.
We take, conservatively,  $5 \times 10^{-6}$ 
as an estimate of the remaining  theoretical uncertainty for values of $x$ away from 
kinematic boundaries.

\begin{figure}[ht]
\centerline{
\psfig{figure=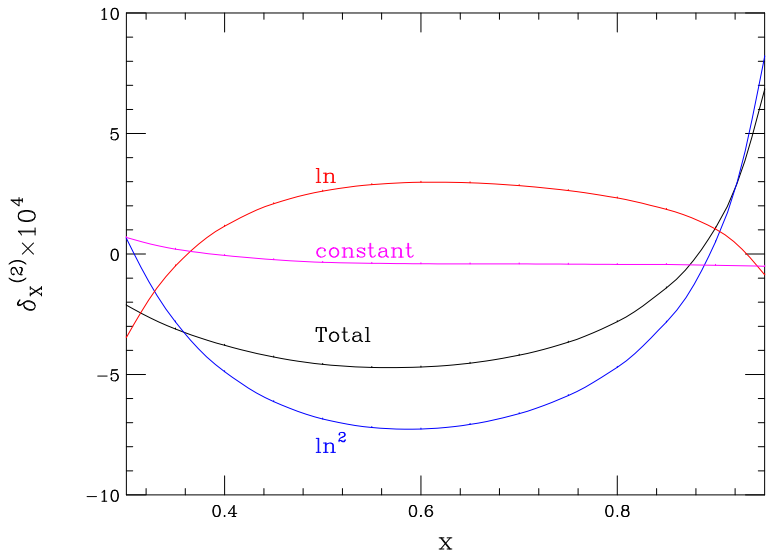,width=4.0in,angle=90}
}
   \caption{\label{OSterms} The ratio of the ${\rm ln}^2$, ${\rm ln}$, and constant NNLO coefficients, as well as the total result, 
     relative to the tree result, 
     $\delta^{(2)}_X=(\alpha / \pi)^2 f^{(2)}_X(x)/f^{(0)}(x)$, versus the electron energy fraction $x$.  The $y$-axis has been
     scaled by $10^4$.
}
\end{figure}

An interesting feature of the electron energy spectrum is that 
effects of radiative corrections are enhanced by large logarithms
of the ratio of the muon mass over the electron mass.  The computation of
the logarithmically enhanced terms in the spectrum is a much 
simpler problem than the full calculation  reported in this paper.
Since large logarithms are routinely exploited 
in theoretical physics for a simplified description, it is interesting 
to gain  some experience on how well this approximation works in various calculations. 
To do so, we compare 
the double- and single-logarithmic 
enhanced corrections with  the full second order QED correction 
to the electron energy spectrum in Fig.~\ref{OSterms}. 
We see that the ratio of the single logarithmic 
term over the constant term obeys the expectation 
$|f^{(2)}_1(x)/ f^{(2)}_0(x)| 
\sim {\rm ln}(m_{\mu}/m_e) \approx 5$, while the ratio of 
the double logarithmic term over the single logarithmic term doesn't: 
$|f^{(2)}_2(x)/ f^{(2)}_1(x)| <5$. Moreover, we note that the
double-logarithmic terms overestimate the full correction.
Because of the cancellation between the doubly- and singly-logarithmic enhanced 
terms, the relative importance of $f^{(2)}_0(x)$ increases. For 
example, at $x=0.5$, the constant term $f^{(2)}_0(x)$ changes 
the second order correction by about $10\%$; this should 
be compared with the naive estimate $1/\ln^2(m_\mu/m_e) \sim 
4 \%$. From this we conclude that the leading logarithmic corrections give a correct order-of-magnitude estimate;
however, the full result can deviate from the leading logarithmic approximation by a factor of $2-3$.

\section{ Conclusions} 

In this paper, we have presented a calculation of the ${\cal O}(\alpha^2)$ 
QED corrections to the electron energy spectrum in muon decay.  The NNLO QED
corrections, relative to the tree level result, are in the range  $-5 \,\, {\rm to} \,\, 8 \times 10^{-4}$, 
depending on the electron energy.  This is larger than the $10^{-4}$ precision expected from 
the TWIST experiment at TRIUMF.  The corrections contain logarithmically enhanced terms of 
the form ${\rm ln}(m_{\mu}/m_e)$, which have been calculated previously in~\cite{Arbuzov:2002pp,Arbuzov:2002cn}.  
The new result derived here is the constant term without logarithmic enhancement, which influences the 
spectrum at the level of $\sim 0.5 \times 10^{-4}$.  The inclusion of this correction reduces 
the theoretical uncertainty below the anticipated experimental precision.  We have argued the the 
remaining uncertainty is at the level of $5 \times 10^{-6}$, which is negligible for any foreseeable experiment. 

Although only the electron energy spectrum is considered in this paper, the computational 
method introduced is flexible enough to permit a computation of {\it any} distribution 
in muon decay, with arbitrary restrictions on the kinematic variables of the electrons and photons.  
The calculation reported here can therefore be extended in several ways.  For the TWIST experiment, there 
are two natural extensions: (1) to include polarization of the muon, which is present in the 
experimental setup; (2) to also constrain the lab-frame angle ${\rm cos} \,\theta$ of the electron, in order to 
match the fiducial region used by the TWIST experiment in their first analysis, $0.50 \leq {\rm cos} \,\theta \leq 0.84$~\cite{twistik}.

There are several applications of our method beyond muon decay, where precise calculations 
of the decay spectra of massive particles are required.  Higher order corrections to semileptonic and radiative $b$ decays are needed 
for extraction of CKM matrix elements and fundamental parameters in heavy quark physics, and in searches for 
new physics.  Decay distributions of top quarks, Higgs bosons, and new massive particles will be precisely measured at the LHC and 
at a future linear collider, and will be used to elucidate the underlying theory describing what is discovered.  We anticipate that 
the techniques developed here will be useful in performing these analyses.

{\bf Acknowledgments.} This research was supported by the
US Department of Energy under contract 
DE-FG03-94ER-40833 and the Outstanding
Junior Investigator Award DE-FG03-94ER-40833, and
by the National Science Foundation under contracts
P420D3620414350, P420D3620434350.


\end{document}